\documentclass[twocolumn]{aastex7}

\shorttitle{processing of silicates} \shortauthors{J\"ager et al.}

\begin{document}

\title{Desorption of water from cosmic silicates}


\author{C. J\"ager}
\affil{Max Planck Institute for Astronomy, Heidelberg, Laboratory
Astrophysics and Cluster Physics Group, Institute of Solid State
Physics, Friedrich Schiller University Jena, Helmholtzweg 3, 07743
Jena} \email{cornelia.jaeger@uni-jena.de}

\author{D. Köhler}
\affil{Analytische Mineralogie, Institute of Geoscience, Friedrich-Schiller-University Jena, Lessingstraße 14, 07745 Jena,
Germany.} \email{}
\and
\affil{Leibniz-Institut für Photonische Technologien e.V. Albert-Einstein-Straße, 07745 Jena, Germany.}

\author{Th. Henning}
\affil{Max Planck Institute for Astronomy, K\"onigstuhl 17, D-69117
Heidelberg} \email{}

\author{F. Langenhorst}
\affil{Analytische Mineralogie, Institute of Geosciences,
Friedrich-Schiller-University Jena, Lessingstraße 14, 07745 Jena,
Germany.} \email{}

\begin{abstract}
Recent JWST and ALMA observations of protoplanetary disks have identified    water reservoirs in the inner regions of the disks. This would provide terrestrial planets forming therein access to a water reservoir. In situ formation of water molecules on grain surfaces was proposed as one possible formation process in the PDS 70 system, but transport by small silicate particles coupled to the gas could be another mechanism. Laboratory experiments on the thermally induced desorption of water molecules from different water-containing silicates at elevated temperatures provide another possibility for the enrichment of water molecules in the inner disk. Physically and chemically bound water in silicates can be released upon annealing processes. Water that is physically bound undergoes desorption at temperatures up to 473\,K. In contrast, chemically bound water in form of different silanol or Mg-OH groups in amorphous silicates or phyllosilicates, can be released at higher temperatures, ranging from 773 to 923\,K. The relevant temperatures are sufficient to trigger the solid-state recrystallization of amorphous silicates to Mg-rich orthopyroxene and olivine. The existence of these crystalline silicates has been substantiated through Spitzer and JWST-MIRI observations. Based on the experimental results, we postulate that the desorption of bound hydroxyl groups may serve as a mechanism for the formation of the water reservoir at temperatures up to 900\,K, much  higher than the water sublimation temperature of 100 to 200\,K.   
\end{abstract}


\section{Introduction}
Silicates and carbonaceous grains are two prevalent components of cosmic dust \citep{Henning:98,Henning:10}. These particles are regarded as precursors and building blocks of planetesimals and eventually planets in the inner regions of protoplanetary disks. In molecular clouds and protoplanetary disks, dust grains are covered by molecular ice layers composed of various molecules, including water, carbon monoxide (CO), carbon dioxide (CO$_2$), ammonia (NH$_3$), methanol (CH$_3$OH), and methane (CH$_4$). The dust grains provide the surface for interstellar ices and create conditions conducive to a rich chemistry in the ice layers and at the interface between ice and dust \citep[for a review]{Boogert:15}. 

Silicates in protoplanetary disks have been found to have a high Mg/Si ratio, and most of the crystalline silicates are found to be iron-poor \citep{Juhasz:10,Sturm:10,Sturm:13,Sargent:09,Olofsson:09}. 
Spitzer and JWST observations have demonstrated that the inner disk region can be water-rich depending on the structure of the disks \citep[e.g.][]{Pontoppidan:10,Banzatti:23,Gasman:25,Banzatti:20,Temmink:25}. It is generally assumed that icy pebbles transport the material across the snowline, where water ice is converted in water vapour. ALMA observations of the young star HL Tau show centrally peaked bright emission from transitions of the main water isotopologue indicating the presence of water within 17 au \citep{Facchini:24}.

JWST observations of the inner disk of PDS\,70 have recently revealed the presence of water vapour in this region \citep{Perotti:23}. PDS\,70 is a protoplanetary disk in the late stage of accretion characterized by a large gap carved by giant planets which reduce the pebble flux. The observed water molecules were found to have temperatures of up to 600\,K. This finding indicates the presence of a water reservoir accessible to potential terrestrial planets forming in this region. The authors suggested that the column densities of water indicate the in-situ formation via a reaction sequence involving O, H$_2$, OH, and survival through water self-shielding. Another possibility could be transport of water bound to small silicate particles well-coupled to the gas. The presence of small silicate particles was already substantiated by Spitzer observations and confirmed with JWST/MIRI spectroscopy. The spectrum exhibits both amorphous and crystalline silicate features \citep{Perotti:23,Liu:25,Jang:24}.

The thermal desorption of water molecules bound to silicate particles is a complex process which depends on the material and the morphology of the grains \citep{Fraser:01,Potapov:18}. Temperature programmed desorption studies of water ice starting around 10\,K show slightly different behavior for different materials, but the desorption process is completed at around 190\,K. It has been experimentally demonstrated that water molecules can be prevented from complete evaporation from the silicate surfaces above the desorption temperature of water ice, thereby enabling their trapping in within the silicates \citep{Zhuravlev:01,Potapov:24}. 

However, water molecules have the capacity to interact with the dust surface, thereby forming different types of bonds. Van der Waals bonds are known for their low bonding energies, and water can be desorbed at low temperatures up to 473\,K \citep{Iler:79,Zhuravlev:01}. Water, when chemically bound, exhibits higher bonding energies, which consequently results in desorption at significantly higher temperatures than around 473\,K \citep{Iler:79}.

In this paper, we investigate a scenario where water could be transported by small silicate grains to higher temperatures as the snowline as a potential explanation for the presence of water in the inner disk regions of PDS\,70. The present study involves desorption experiments of various magnesium silicates, each containing different amounts of physisorbed and chemisorbed water. Laboratory silicate dust analogs, as well as natural minerals, were subjected to thermogravimetric analyses. These analyses were used to illustrate the possible enrichment of warmer inner dust regions with water molecules from silicates.    

\section{Experimental}

The study is focused on the desorption behavior of magnesium silicates containing internal water. Various samples were examined, including sol-gel synthesized samples and natural minerals such as serpentine minerals, which are known to be important components in carbonaceous chondrites and the recently returned Ryugu particles \citep{Leroux:24}, shown in Table\,\ref{mat}.  Amorphous magnesium silicates of various stoichiometries were produced by the sol-gel technique. The sol-gel process is based on the chemical polymerization of silicates in a liquid phase at low temperatures. Metal organic compounds such as tetramethoxysiloxane (TMOS) and magnesium methylate Mg(OCH$_3$)$_2$ served as precursors in the synthesis of the desired compounds. The metal organic compounds were dissolved in water and ethanol. Silicate formation is a two-step process. The initial process entails the hydrolysis of the precursors, which results in the formation of hydroxides. Subsequent to this, the hydroxides undergo a condensation process, yielding a three-dimensional magnesium silicate network composed of Si(OH)$_4$ and Mg(OH)$_2$. Hydrolysis and condensation are chemical equilibrium reactions that proceed in a competitive manner. In order to avoid phase separations and the formation of inhomogeneous silicate material, it is imperative to select silicon and magnesium components that exhibit nearly equivalent hydrolysis and condensation rates.

\begin{table*}[tb] \caption{The silicate samples employed for thermogravimetric studies.}
\begin{center}
\begin{tabular}{lcc}
\hline
                                                                             Formula/Name             & Mineral name     & Origin             \\               \hline
Laboratory synthesized analogs                                                                        &                  &                  \\  
Mg$_{0.7}$SiO$_{2.7}$                                                                                 &amorphous silicate& Sol Gel Synthesis\\              
Mg$_2$SiO$_4$                                                                                         &amorphous silicate& Sol Gel Synthesis\\
Natural minerals                                                                                      &                  &                  \\
(Mg$^{2+}_{5,471}$Fe$^{2+}_{0,432}$Al$^{3+}_{0,069}$Si$^{4+}_{3,945}$Al$^{3+}_{0,069}$O$_{10}$(OH)$_8$& antigorite (serpentine mineral) & Erbendorf/Oberpfalz\\ 
(Mg,Fe,Ni)$_6$Si$_4$O$_{10}$(OH)$_8$       & antigorite and lizardite (serpentine minerals)         & Zöblitz\\
\hline
\end{tabular}
\label{mat}
\end{center}
\end{table*}

The synthesis of the amorphous magnesium silicates Mg$_{0.7}$SiO$_{2.7}$ and Mg$_2$SiO$_4$ was realized through an advanced modification of the H$_2$O$_2$-assisted sol-gel reaction of magnesium methylate (Mg(OCH$_3$)$_2$) and TMOS  based on the synthesis proposed by \citet{Burlitch:91} \citep[see also][]{Agladze:96}. The hydrogen peroxide serves as a proton provider and a catalyst. The synthesis must be carried out in a purified Ar atmosphere to prevent the precipitation of insoluble Mg(OH)(OCH$_3$) and the phase separation of MgO.

\begin{table} \caption{Composition of the amorphous magnesium silicate samples  measured by EDX analyses in the TEM.}
\label{compo}
\begin{tabular}{ccc}
\hline
Sample name           & MgO        & SiO$_2$         \\
                      & \multicolumn{2}{c}{wt \%}    \\
\hline
Mg$_{0.7}$SiO$_{2.7}$ &  31.2     &  67.8        \\
Mg$_{2}$SiO$_{4}$     &  56.6     &  42.4        \\
\hline
\end{tabular}
\label{EDX_SG}
\end{table}

Following the removal of the solvents methanol and water from the produced gels at temperatures up to approximately 473\,K, the magnesium silicate powder was subjected to heating at temperatures between 773 and 873\,K. This process was intended to remove the residual water, thereby achieving a densification of the silicate framework and decreasing the porosity of the grains. The sample of Mg$_{0.7}$SiO$_{2.7}$ was subjected to annealing at a temperature of 773\,K for a duration of one hour. Mg$_2$SiO$_4$ was annealed for a period of 3 hours at a temperature of 793\,K. The composition and phase homogeneity, as well as the densification process, were subsequently examined via high-resolution transmission electron microscopy (JEOL JEM 3010, acceleration voltage of 300\,kV) equipped with an energy dispersive X-ray (EDX) analytical system. The detailed EDX analyses of the amorphous silicates are presented in Table\,\ref{EDX_SG}. The error of the quantitative EDX analysis of the given composition was estimated with $\pm$0.3 at\% based on numerous measurements and comparisons with standard materials.

The studied serpentinite sample was obtained from the vicinity of the village Erbendorf in the Upper Palatinate and is predominantly composed of lamellar antigorite, accompanied by a trace of magnetite. Antigorite, a lamellated, monoclinic mineral in the phyllosilicate serpentine subgroup with a standard chemical formula of (Mg,Fe$^{2+}$)$_3$Si$_2$O$_5$(OH)$_4$ was found to be the main phase.  Antigorite contains a high fraction of water of up to 13 wt\% \citep{Dodony:02,Ulmer:95}. The second sample is a serpentinite from the village Zöblitz (Erzgebirge) which is mostly made of antigorite and lizardite, both of the same stoichiometry but slightly different sheet structure.

Thermogravimetric analyses have been performed in a temperature range between 298 and 1273\,K with a heating rate of 20\,K\,min$^{-1}$ at a pressure of 1$\times$10$^{-3}$ mbar using a NETZSCH TG 209 F1 Libra instrument. The powdered siliates were heated up in Al$_2$O$_3$ crucibles. The accuracy of the desorption temperatures determined in the thermogravimetric analyses was estimated to be within $\pm$3\,K. This encompasses the error associated with the temperature measurement and the error related to the annealing rate.

In addition, we performed directly coupled evolved gas analyses (DEGAS) by combining thermogravimetry (STA 429 Netzsch) with quadrupole mass spectrometry (QMA-125 Balzers) between 298 and 1673\,K. The thermogravimetric analyses have been conducted at a pressure of 10$^{-6}$\,mbar.  The technique aimed to identify the release of water from the phyllosilicates at different heating rates of 10, 5, and 1\,K\,min$^{-1}$.  

The powdered materials were used to prepare KBr pellets for IR transmission measurements. Spectroscopy was performed using a BRUKER 113v FTIR spectrometer in the 2--25 $\mu$m range, with a DTGS/KBr detector in combination with a KBr beamsplitter. The applied spectral resolution was 2 cm$^{-1}$ (R=1800).

\section{Silicate Dust Materials}

In crystalline pyroxenes Mg$_{x}$Fe$_{1-x}$SiO$_3$ and olivines Mg$_{2x}$Fe$_{2-2x}$SiO$_4$ (x=1), the silicate structure is well defined. Pyroxenes consist of SiO$_4$ chains, which are linked by the  divalent metal ions Mg$^{2+}$ and Fe$^{2+}$. In olivines, the SiO$_4$ tetrahedra are isolated and again linked by the  cations. In both materials, the divalent cations Mg$^{2+}$ or Fe$^{2+}$ are octahedrally coordinated by six oxygens. Mg$^{2+}$ and Fe$^{2+}$ can replace each other in the olivines and pyroxenes forming solid solutions of either MgSiO$_3$--FeSiO$_3$ or Mg$_2$SiO$_4$--Fe$_2$SiO$_4$. The mineralogical endmembers are denoted as enstatite and ferrosilite, or forsterite and fayalite, respectively. 

In contrast, the short-range and long-range structure of amorphous silicates of pyroxene  or olivine  stoichiometry is not well defined. However, SiO$_4$ tetrahedra remain the main building blocks of the amorphous pyroxene- and olivine-type  silicates \citep[for a review]{Henning:10}, \citep{Jaeger:03,Kohara:11,Goumans:11}. The incorporation of Mg or Fe into a network of linked SiO$_4$ tetrahedra leads to the destruction of Si-O-Si bridges between the tetrahedra resulting in the formation of nonbridging oxygen. In the group of cosmic silicates with pyroxene and olivine stoichiometry, the iron-free silicates are preferentially observed.  The amorphous structure in the magnesium silicates is characterized by SiO$_4$ tetrahedra and MgO$_x$ polyhedra. The coordination numbers of Mg demonstrates significant variability \citep{Kohara:04,Kohara:11} and does not coincide with the coordination number six in the well-defined structures of crystalline silicates. 

Olivines and pyroxenes can be hydrated and transformed into serpentine minerals, a process called serpentinization \citep{Moody:76,Beard:07}. It is a consequence of the hydration and oxidation of olivines and pyroxenes. During this process, significant quantities of water can be inserted into the minerals. Serpentinites are metamorphic rocks that contain the serpentine minerals antigorite, lizardite, and chrysotile. Other hydroxyl-bearing minerals forming during serpentinization are brucite and talc \citep{Moody:76,Dodony:02}. Forsterite has the potential to transform into serpentine (Mg$_3$Si$_2$O$_5$(OH)$_4$) and brucite (Mg(OH)$_2$), whereas enstatite can be transformed into talc (Mg$_3$Si$_4$O$_{10}$(OH)$_2$) or serpentine. 
Serpentines are phyllosilicates with a layered structure consisting of a tetrahedral (Si$_2$O$_5^{2-}$) sheet and an octahedral (Mg$_3$(OH)$_6^{4+}$) sheet.
Every serpentine mineral has its distinct layer structure \citep{Morris:76}. 

\begin{figure}[tb]
\epsscale{1.0}
\plotone{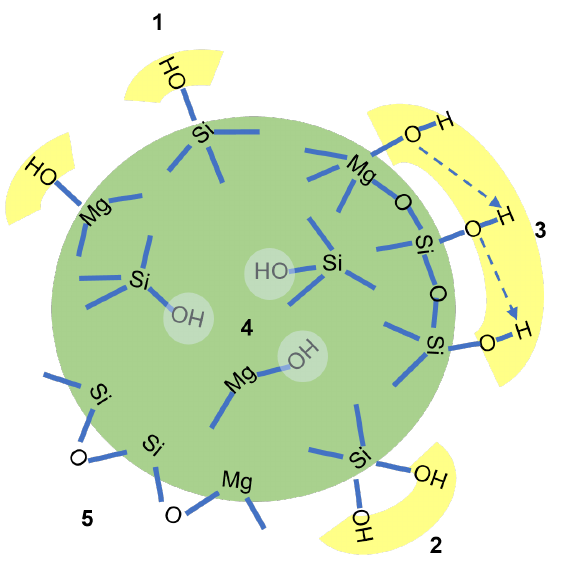}
\caption{Possible structural characteristics of chemisorbed water in amorphous magnesium silicates. It can be distinguished between isolated (1), geminal (2), vicinal (3), and internal hydroxyl groups in amorphous magnesium silicate. The dashed lines represent hydrogen bonds between adjacent surface OH groups. After the removal of chemisorbed OH, typical siloxan or bridging oxygen groups remain on the surface or internally.\label{hydroxyl}}
\end{figure}

Both, the sol-gel produced amorphous silicates and the serpentinite minerals are characterized by a certain amount of OH groups and water. These OH groups and water molecules can bind to amorphous silicates in different ways, as illustrated in Fig.\,\ref{hydroxyl}. 

In amorphous magnesium silicates, surface OH groups bound to silicon or Mg are subdivided into different groups: single silanol (Si-OH) or Mg-OH (magnesium hydroxyl) groups; geminal silanol Si(OH)$_2$ groups; and vicinal or bridged OH groups. OH-free Si--O--Si groups called siloxane groups or Mg--O--Si are also available on amorphous silicates \citep{Iler:79,Zhuravlev:01}. In addition, there is structurally bound water inside the magnesium silicate structure called internal silanol or Mg-OH groups.

\begin{figure}[tb]
\epsscale{1.2}
\plotone{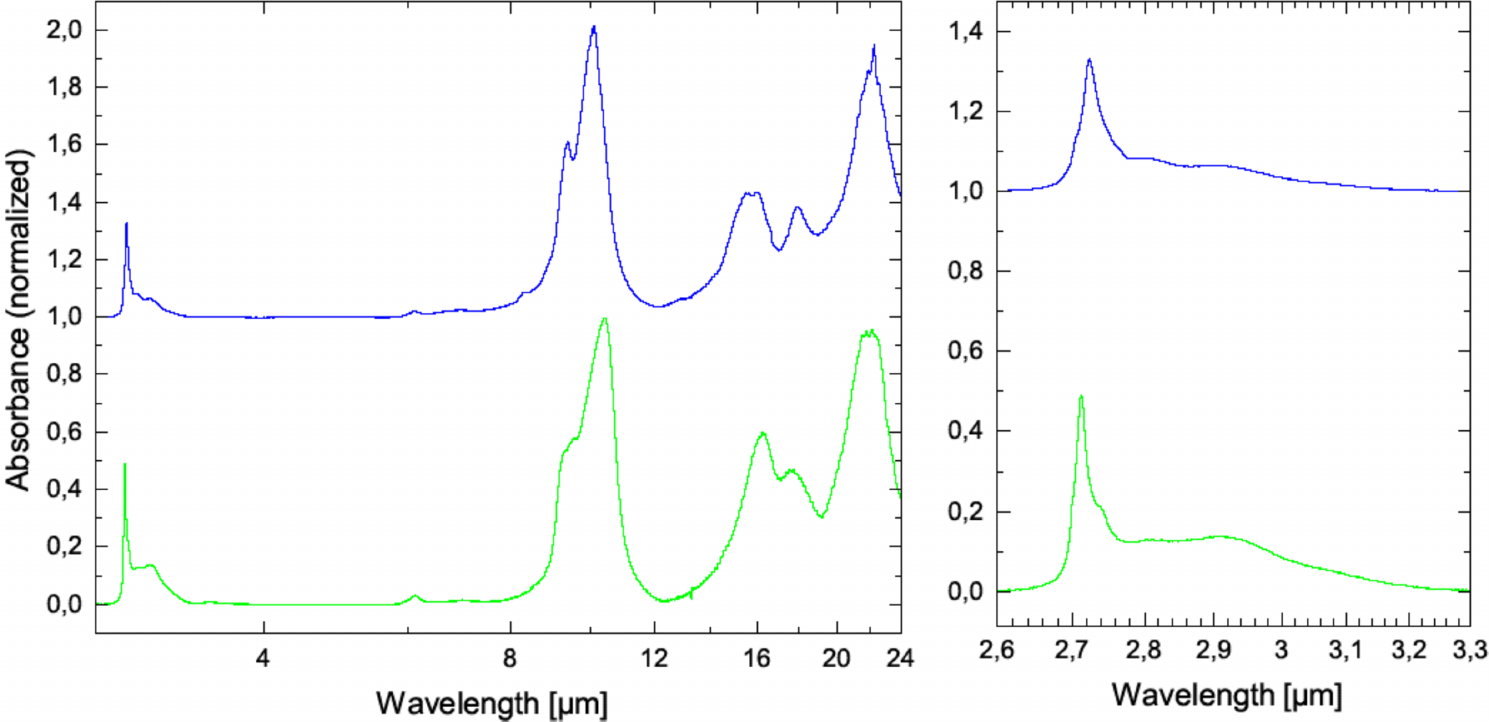}
\caption{IR spectra of the two serpentinite materials. The lower spectrum (green curve) is attributed to the serpentinite from Zöblitz, while the upper spectrum (blue) corresponds to the serpentinite from Erbendorf. The curve has been shifted vertically to enhance visibility. The right graph depicts the range of the OH vibrations.\label{IR_serp}}
\end{figure}

\begin{figure}[tb]
\epsscale{1.2}
\plotone{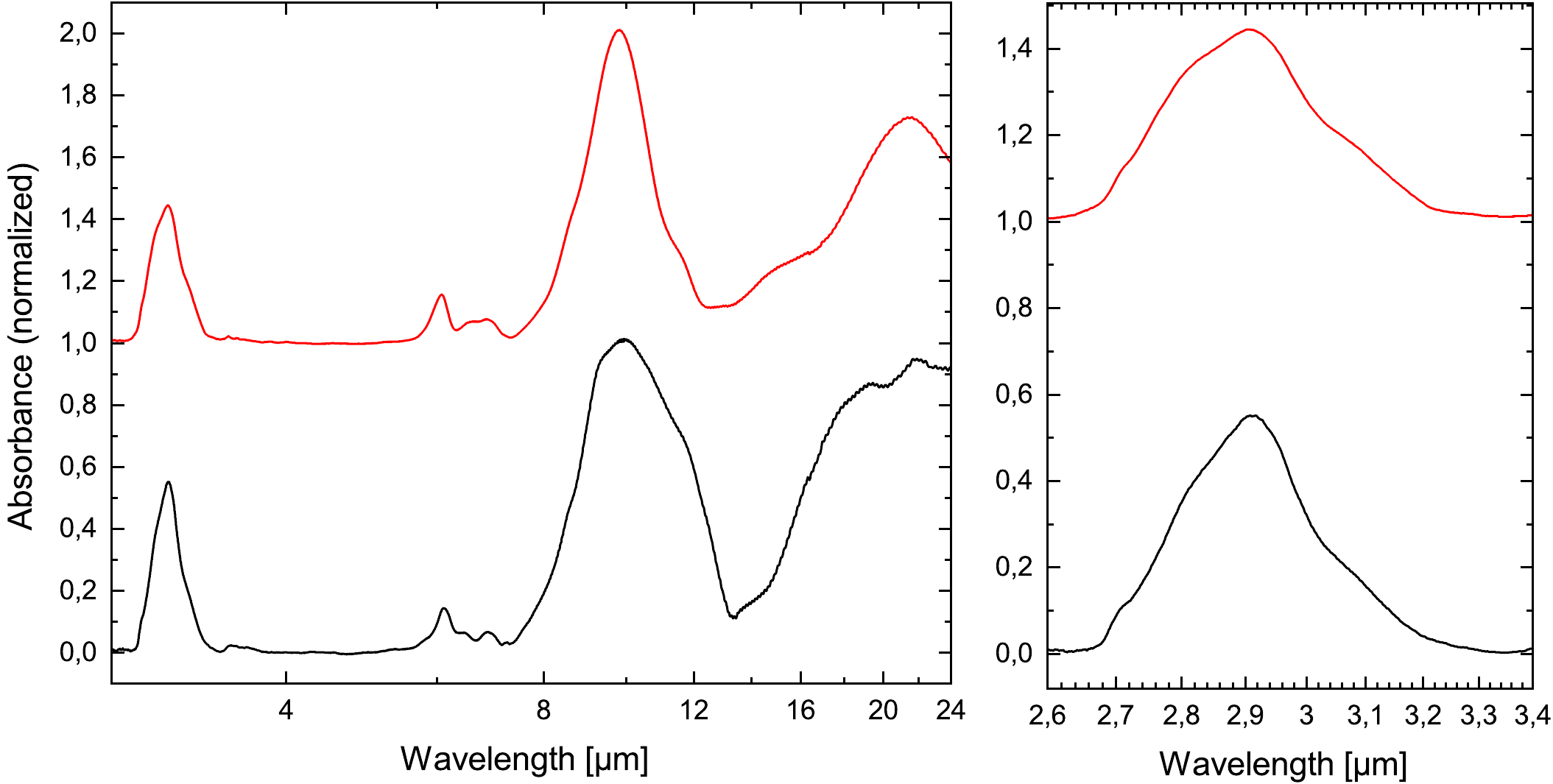}
\caption{The left-hand graph illustrates the IR spectra of the water-containing amorphous silicates. The upper spectrum(red curve) corresponds to Mg$_{0.7}$SiO$_{2.7}$. The lower spectrum (black curve) represent Mg$_2$SiO$_4$. The upper curve has been shifted to enhance clarity. The right graph indicates the range of OH vibrations of the minerals before annealing.\label{IR_Solgel}}
\end{figure}

The IR spectra of the water containing sol-gel silicates and serpentinite minerals are displayed in Figs.\,\ref{IR_serp} and \ref{IR_Solgel}. 

The serpentinite spectra reveal the characteristic narrow, crystalline silicate features of serpentine \citep{Wu:21}. In the range between 8 and 12.5 $\mu$m, asymmetric and symmetric Si--O stretching modes are visible. The series of bands in the 20--25 $\mu$m range is attributable to the bending of Si–O bonds and/or the stretching of Mg–O and Fe–O bonds in the silicates \citep{Fritsch:21}. 
The stretching vibrations of Si–O and other Me-O bonds, in addition to the bending vibrations, have been observed to vary considerably among different serpentinite minerals. In addition, the characteristic OH stretching vibrations are visible between 2.6 and 3.1 $\mu$m. The sharp features at around 2.7 to 2.8 are usually attributed to OH-stretching vibrations on the outer surface and outer sloping stretching OH groups of the serpentine mineral \citep{Wu:21, Kloprogge:99}.
The bands observed at longer wavelengths are also attributable to OH bands in  crystalline silicates that are bound to specific sites and surfaces \citep{Galuskin:07}.   

The amorphous silicates show the typical broad silicate features at 10 and 20 $\mu$m associated with the Si--O--Si and Si-OH stretching and bending modes and vibrations of Mg--O, respectively, and broad OH stretching bands in the range between 2.7 and 3.1 $\mu$m. The OH bands at around 3.0 $\mu$m are more complex compared to those of crystalline silicates because of a distribution of bond lengths and angles and less clearly defined OH positions. These features can be ascribed to single and predominantly paired vicinal and geminal OH groups accompanied by contributions from physisorbed molecular water at the long wavelength tail of the band \citep{Iler:79,Bertoluzza:82}. Contribution to the vibrational band at around 2.9 $\mu$m and higher can also be attributed to OH groups bound to Mg in tetrahedral coordination \citep{Suzuki:00} demonstrating the complex structural situation of statistically disorederd OH groups in amorphous silicates. 

The strength of the OH bands in amorphous and crystalline silicates correlates with the amount of water incorporated, however it requires an exact calibration for the corresponding compositions, structures, and OH content \citep{Libowitzky:97,Behrens:03}.   

\section{Results} 
\begin{figure}[tb]
\epsscale{1.2}
\plotone{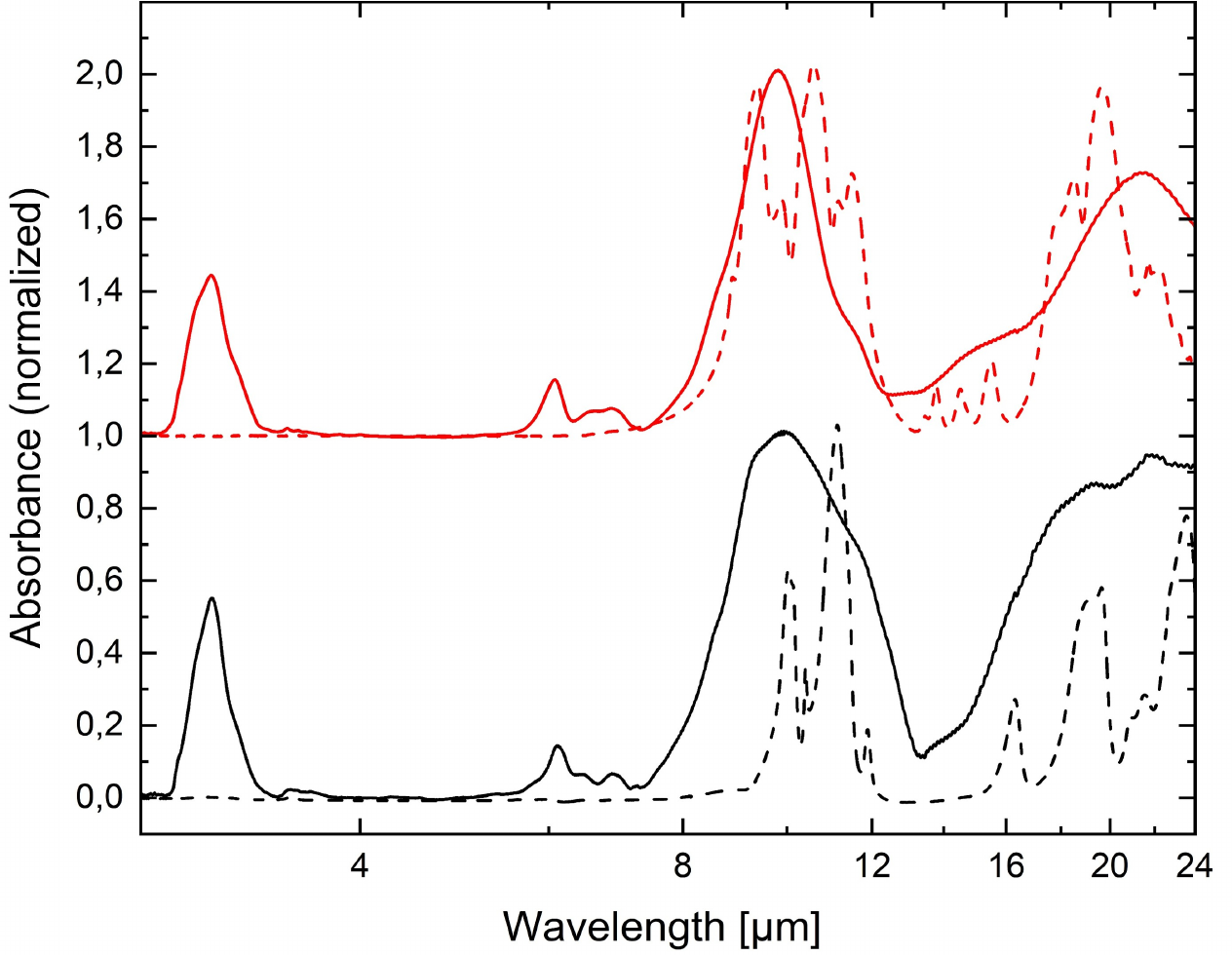}
\caption{IR spectra of the water-containing amorphous silicates in comparison to the annealed samples. The upper spectra(red solid and dashed curve) correspond to Mg$_{0.7}$SiO$_{2.7}$ before and after annealing, respectively. The lower spectra (black solid and dashed curve) represent Mg$_2$SiO$_4$ before and after annealing. The Mg$_{0.7}$SiO$_{2.7}$ curves have been shifted for better visibility.  \label{IR_Solgel_anneal}}
\end{figure}

The thermal treatment of the various silicates results in the desorption of water at distinct temperature ranges. These changes can be observed in the infrared (IR) spectral range, as illustrated by a loss of the spectral OH stretching and bending vibrations in all samples. As shown in Fig.\,\ref{IR_Solgel_anneal}, the original and the corresponding annealed samples of the amorphous sol-gel samples are shown. The Mg$_{0.7}$SiO$_{2.7}$ sample exhibits typical enstatite infrared (IR) patterns upon annealing, while Mg$_2$SiO$_4$ is converted into forsterite. A comparable behaviour can be observed in the serpentinite material, which undergoes a loss of OH vibrational bands. Mixtures of enstatite and forsterite with their typical crystalline IR structures are formed.

\begin{figure}[tb]
\epsscale{1.4}
\plotone{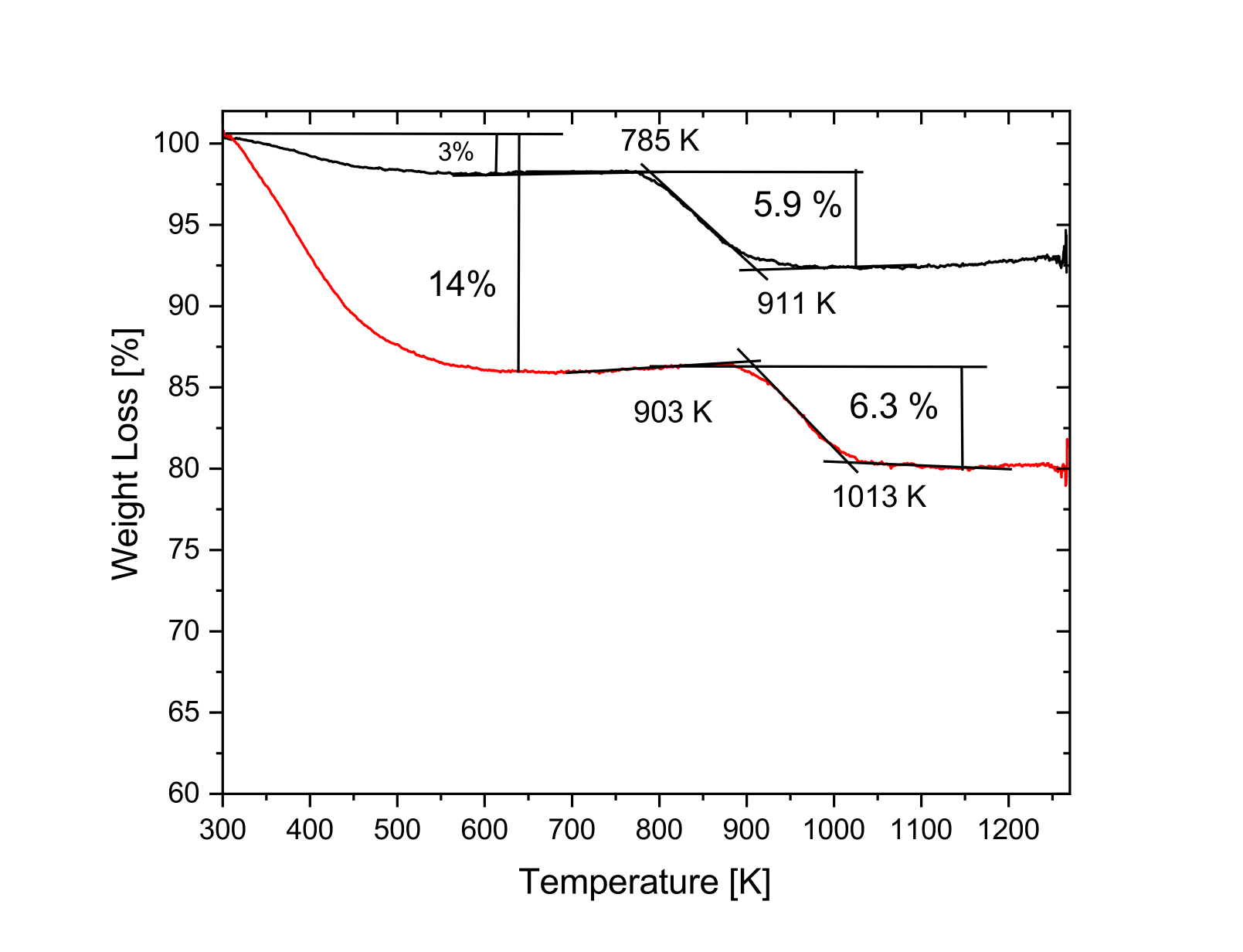}
\caption{Thermogravimetric analyses showing the desorption behavior of sol-gel-synthesized amorphous  silicates with the compositions Mg$_{0.7}$SiO$_{2.7}$ (red curve) and Mg$_2$SiO$_4$ (black curve) .\label{Tg_SG}}
\end{figure}

The desorption of water at distinct temperature ranges induces a significant mass loss. This phenomenon can be measured and quantified through thermogravimetric analysis. 
As illustrated in Fig.\,\ref{Tg_SG}, the mass loss behavior of the amorphous silicates produced by the sol-gel technique is observed under annealing conditions up to 1273 K.

A slightly distinct mass loss behavior is evident between the two silicates. The Mg$_{0.7}$SiO$_{2.7}$ sample demonstrates a substantial mass reduction within the temperature range up to approximately 473 K. This phenomenon is characteristic of trapped or physically bound water. It has been determined that approximately 14\% of the sample mass is lost during this step. A subsequent stage of water desorption is evident at temperatures ranging from 903 to 1013 K, accompanied by a mass loss of approximately 6.3\%, attributable to the desorption of chemically bound water. 

The amorphous silicate sample with olivine stoichiometry (Mg$_2$SiO$_4$), however, exhibits a slightly divergent behavior. The substance contains a small amount of physically bound water, amounting to 3\% of its total mass. The second desorption step of chemically bound water can be observed at temperatures ranging from approximately 785 up to 911 K, where a mass loss of 5.9\% is evident.
The difference in the low-temperature range for the two sol-gel samples is attributable to the distinct pretreatment and drying procedures prior to the annealing step. The specimen of Mg$_{0.7}$SiO$_{2.7}$ was exposed to temperatures of approximately 493 K for a duration of two hours. This treatment resulted in incomplete removal of physically bound water. The Mg$_2$SiO$_4$ specimen underwent a drying process that lasted approximately six hours. This resulted in a negligible amount of residual physically adsorbed water of approximately 3\%.

In both samples, the water desorption at higher temperatures can be explained by differently structured silicate networks containing OH groups. Vicinal bridged OH groups have been found to disappear around 673 K, while single and geminal OH groups remain on the silica surface \citep{Iler:79}. The removal of internal OH groups occurs at a slightly higher temperature. In both samples, a second mass loss step commences at approximately 773 K and 873 K, respectively. This process can be explained by the removal of vicinal, geminal, and isolated hydroxyl groups from the silicates. The elevated desorption temperature of Mg$_{0.7}$SiO$_{2.7}$ is likely attributable to the presence of a greater number of geminal and isolated hydroxyl groups, as well as a moderately higher desorption temperature of Si-OH groups in comparison to Mg-OH. During the dehydration process, the concentration of siloxane Si-O-Si bridges increases considerably. However, OH groups' occurrence is not limited to the surface of grains; they can also be formed by the dissociation of oxygen bridges between the SiO$_4$ tetrahedra or other metal polyhedra within the silicate skeleton.   

\begin{figure}[tb]
\epsscale{1.4}
\plotone{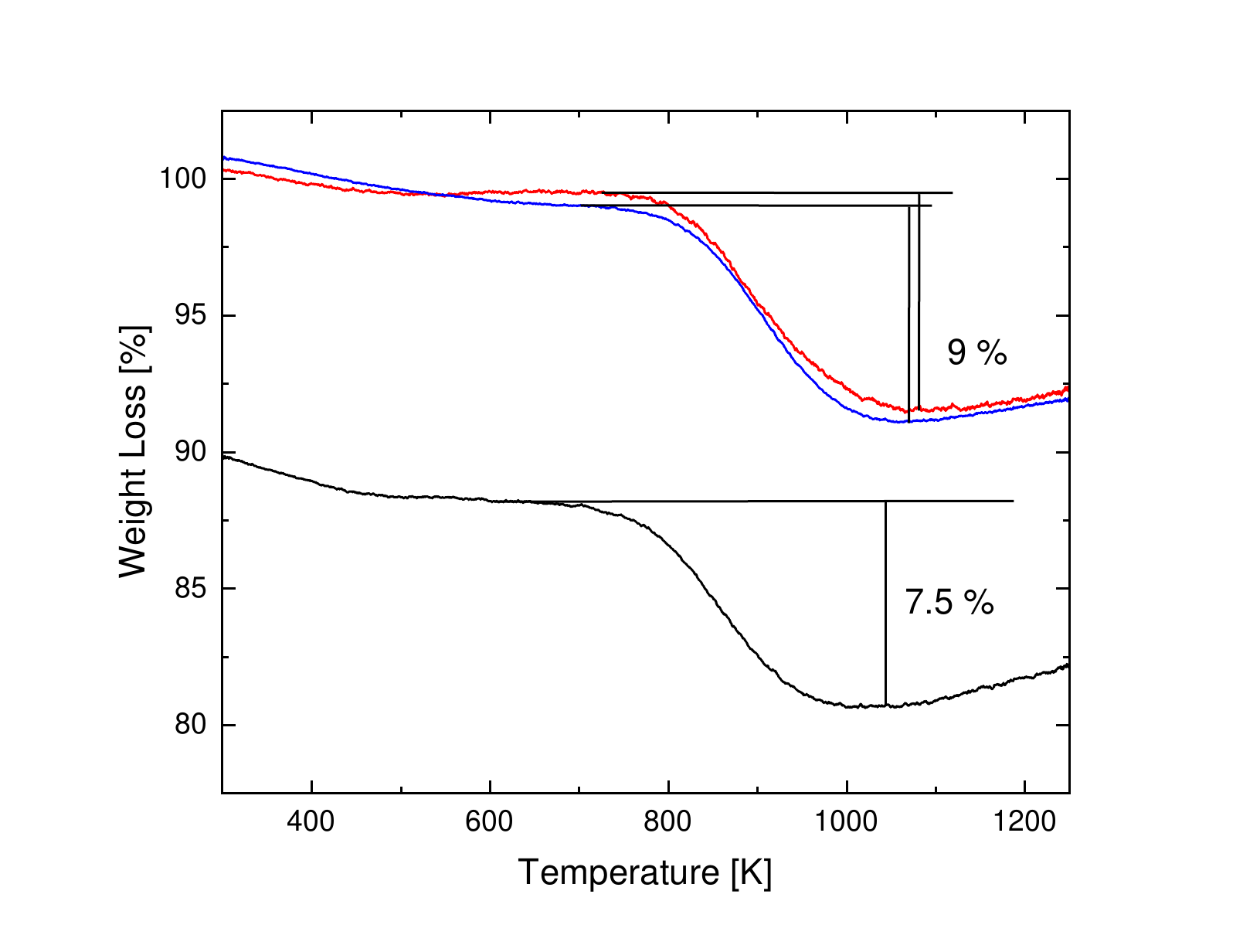}
\caption{The upper two curves represent the mineral serpentine antigorite from Erbendorf recorded at ambient pressure (blue) and 10$^{-3}$ mbar (red). The black curve, which has been shifted by -10\% to facilitate a more comprehensive overview, illustrates the serpentinite from Zöblitz.\label{Tg_min}}
\end{figure}

\begin{figure}[tb]
\epsscale{1.4}
\plotone{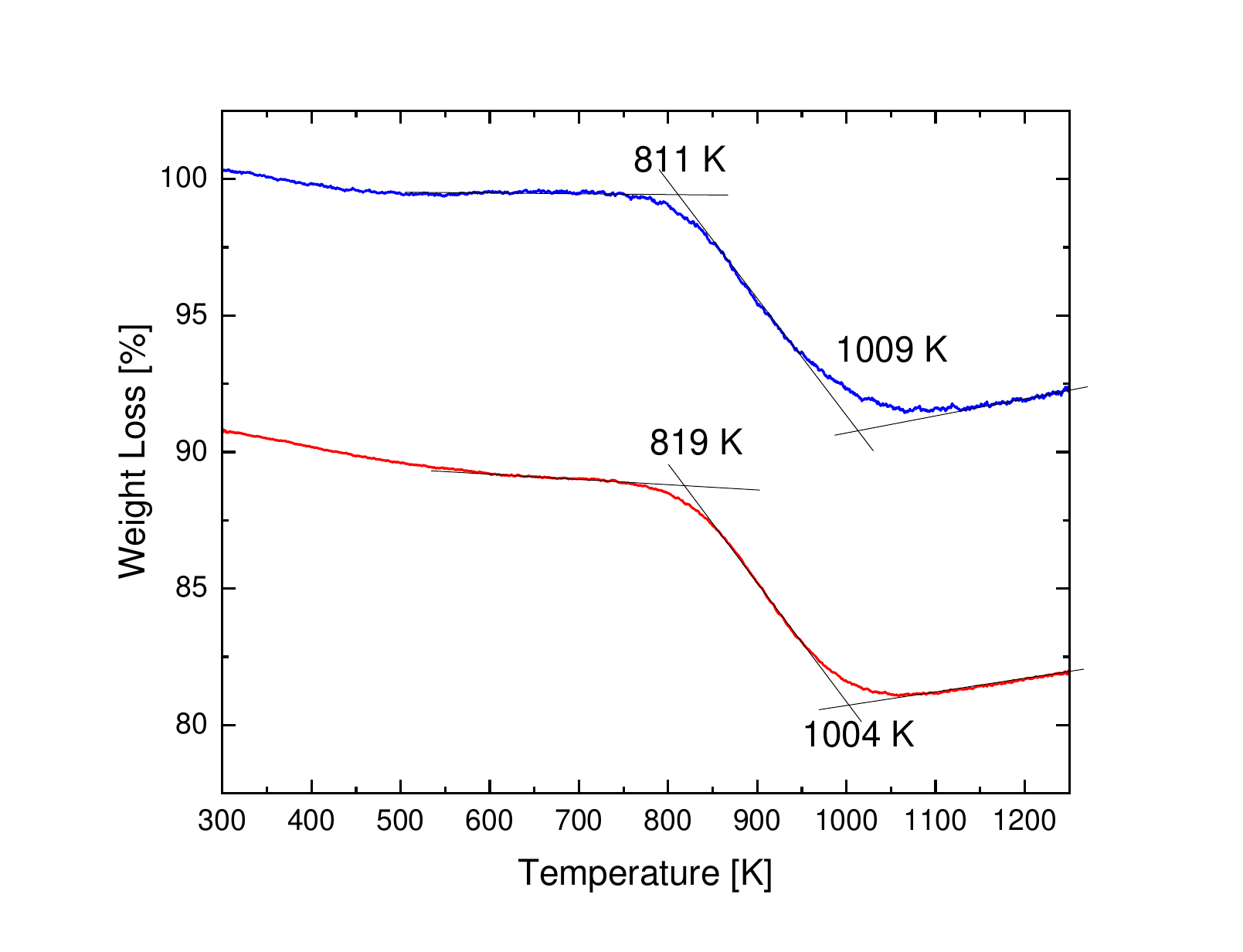}
\caption{A comparison of the desorption temperatures determined for the Serpentinite Erbendorf measured at  ambient pressure (blue curve) and 1$\times$10$^{-3}$ mbar (red curve, shifted by -10\%).\label{Tg_vgl}}
\end{figure}

Fig.\,\ref{Tg_min} presents the desorption behavior exhibited by two distinct serpentine samples. A comparison of the two materials reveals a high degree of similarity in their temperature range of desorption. Following the loss of approximately 1--1.5 \% mass within the temperature range of 293 K to 523 K for the Erbendorf sample, a mass loss of 9 \% was measured between 819 and 1004 K. The thermogravimetric analyses conducted at varying pessures exhibit a highly comparable desorption behavior. As demonstrated in Fig.\,\ref{Tg_vgl}, the onset and end times of the process are the same across both pressure conditions. The measured desorption profiles are comparable to those previously measured for different serpentines \citep{Pohl:24}. 

\begin{figure}[tb]
\epsscale{1.2}
\plotone{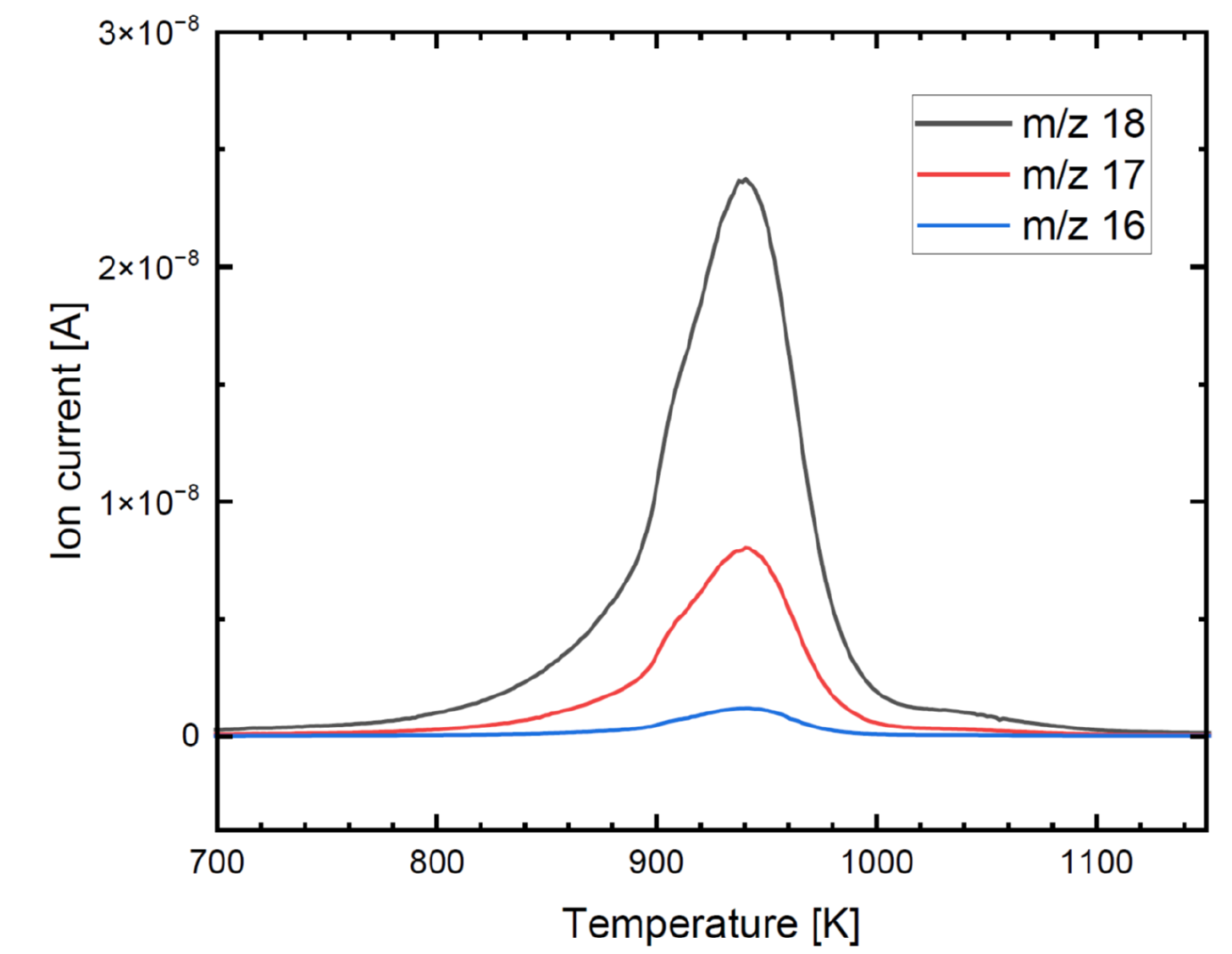}
\caption{Measured ion currents for the masses (m/z) = 18, 17, and 16 as a function of temperature for a heating rate of 10 K min$^{-1}$  confirming a mass loss of water. The m/z = 17 and 16 are expected fragments of the water molecule (m/z = 18) produced by the ionization process in the mass spectrometer.\label{MS_1}}
\end{figure}

\begin{figure}[tb]
\epsscale{1.2}
\plotone{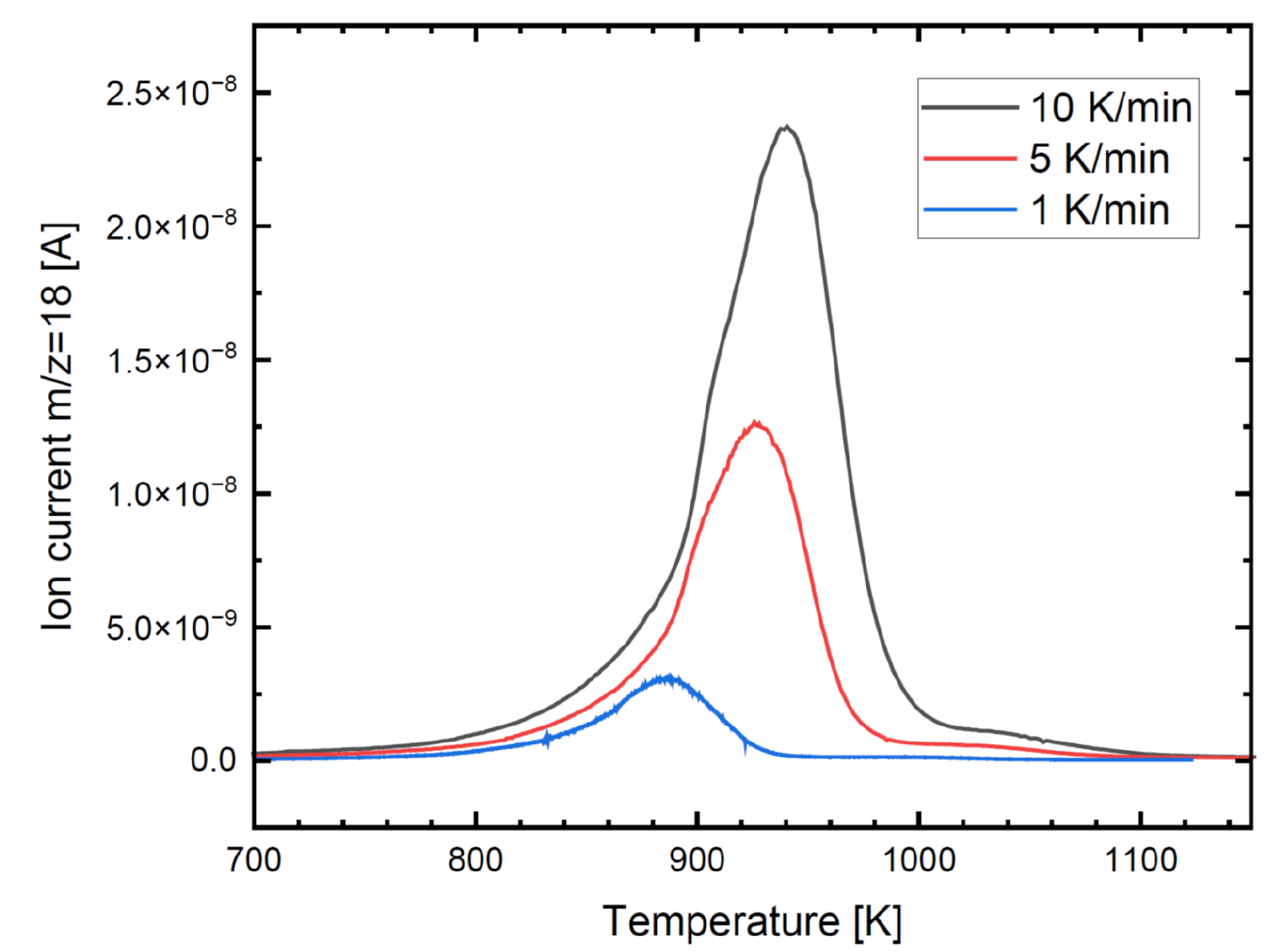}
\caption{DEGAS data of the Erbendorf serpentinite for m/z = 18 (H$_2$O) at three different heating rates (1 K, 5 K, and 10 K\,min$^{-1}$). The peak maxima shift clearly to higher temperature for higher heating rate.
\label{MS_2}}
\end{figure}

To verify that the mass loss is indeed due to the release of water, we have additionally performed DEGAS analy\-ses on the Erbendorf serpentinite at variable heating rates of 1, 5, and 10\,K\,min$^{-1}$ (Fig.\,\ref{MS_2}). These measurements provide mass spectra of the gases and fragments thereof released during heating. At a heating rate of 10 K, we observe an increase of the ion currents for masses per charge (m/z) = 18 (H$_2$O), 17 (OH), and 16 (O) at about 800 K (Fig.\,\ref{MS_1}), coinciding well with the desorption temperature determined by thermogravimetry Fig.\,\ref{Tg_vgl}. Thus, we can confirm that the mass loss beyond this temperature is due to the release of water.

The desorption temperature is determined not only by the specific internal structure of the substance, but also by the heating rate during thermogravimetric analysis.  It has been demonstrated that an increase in the heating rate can lead to an increase in the onset of desorption temperature. \citet{Fofanov:25} have measured the water desorption from SiO$_2$ powder produced by the sol-gel technique. A silanol desorption peak has been observed at 824 K for a heating rate of 10 K\,min$^{-1}$. Thermogravimetry indicates that an increase to 20 K\,min$^{-1}$ has resulted in a shift of the desorption peak by 15 K. DEGAS measurements corroborate this tendency and show that the peak temperature shifts to 886 K, 927 K, and 940 K for the heating rates of 1 K, 5 K, and 10 K\,min$^{-1}$, respectively (Fig.\,\ref{MS_2}). It is possible to derive the activation energy from thermogravimetric measurements or other temperature-programmed experiments measured at different heating rates by using the Kissinger equation. The classical Kissinger equation ln$(\beta/T_p^2) = -E_a/RT_p + C$ can be used to estimate the activation energy of desorption processes measured at different heating rates $\beta=dT/dt$. Plotting ln$(\beta/T_p^2)$ against $1/T_p$ results in a straight line with a slope equal to the activation energy. The determination of the activation energy peak positions facilitates the estimation of the relevant peak temperatures for different annealing rates. The peak values derived at three distinct heating rates were found to correspond to an activation energy of 273\,kJ\,mol$^{-1}$. However, it is imperative to bear in mind that the Kissinger equation constitutes a simplistic approach to the estimation of activation energy.

The high-temperature desorption of water from serpentines or amorphous magnesium silicates cannot generally be treated as a simple surface recombination/desorption problem. In contrast, the prevailing interpretation is that of a bulk solid undergoing dehydroxylation, wherein the observed T$_g$ peak position is determined by factors such as structural transformation, nucleation/growth, interface motion, or diffusion through the dehydroxylated layer.  
Recent analyses of serpentine dihydroxylation interpret the process in terms of solid-state reaction models and phase-transformation mechanisms rather than a surface recombination law. Detailed T$_g$ studies of serpentine dehydroxylation generally fit the kinetics with diffusion- or interface-controlled solid-state reaction models rather than second-order recombinative desorption \citep{Aliza:11}. However, the development of kinetic theories to derive activation energies of water desorption from bulk materials was not the objective of the present study. 


\section{Astrophysical Discussion}\label{Astro}
Recently, the presence of water molecules was identified in the JWST MIRI spectrum of the inner protoplanetary disk PDS\,70, the zone where terrestrial planets are formed \citep{Perotti:23}. This is the first detection of water in a disk that hosts already protoplanets. However, it is not yet understood if the water was formed in situ from O, H$_2$ or OH, or if it was delivered by water-containing dust grains from the outer to the inner disk, where it can desorb and form water vapor. 

In addition to water molecules, silicate dust grains have been detected in the Spitzer IRS and JWST MIRI spectra. The observed 10 and 20 $\mu$m bands have been fitted by a mixture of amorphous silicates with pyroxene- and olivine stoichiometry, as well as SiO$_2$ \citep{Liu:25,Jang:24}. The two crystalline iron-free components, enstatite and forsterite, were also identified. \citet{Perotti:23} proposed that the dust continuum is reproduced with a three-component disk model, with a surface layer of 400--600 K accounting for the bulk of the observed emission. The presence of the two crystalline components is indicative of thermal processing of the silicates.

\citet{Perotti:23} argued that the water molecules might have been formed directly on the surface of the grains from O atoms, hydrogen, and OH groups. 
In the case of PDS\,70, the pebble transport is blocked by two giant planets, but small grains could still cross the gap being well coupled to the gas.   

Complex silicates with amorphous and crystalline structures have the capacity to store a certain amount of water, which is either physically or chemically bound to metal ions in the structure. The hydroxylation of magnesium silicates can be initiated by reactions between Si-O-Si and Mg-O-Mg groups located at the surface of amorphous silicate and the water ice layer, under conditions of low temperature \citep{Thi:20}. In addition, the activation of this process can be triggered by either UV photons or electrons, which have the capacity to effectively dissociate bonds within water ice or dust surfaces \citep{Yu:16,Monti:25}. Moreover, previous experiments have studied the formation of amorphous silicates from molecular species and radicals at temperatures between 10 and 20\,K whithin ices that were contaminated with water molecules. The formed amorphous silicates with pyroxene and olivine composition have been found to contain OH groups, which were detectable in the IR spectrum of the silicates warmed up to room temperature \citep{Rouille:15}. 

\citet{Mennella:20} has demonstrated that the interaction of hydrogen atoms with Mg-rich amorphous silicates also leads to the formation of hydroxyl groups, a process that may occur in the ISM. In addition, the bombardment of amorphous silicates with low-energy protons of a few keV can also result in the formation of chemically bound hydroxyl groups \citep{Djouadi:11}. 

Thermogravimetric studies have demonstrated that the desorption of physically adsorbed water occurs up to a temperature of approximately 473 K. In contrast, chemically bound water is desorbed at higher temperatures, with an onset at around 773 K. This temperature range is consistent with the temperature required for the crystallization of amorphous grains \citep{Jaeger:03,Fabian:00}. The crystallization temperature of amorphous silicates is dependent on the softening range and transformation point, which in turn are influenced by the thermal history of the amorphous material \citep{Vogel:92}. Previous experimental studies on amorphous magnesium silicates have found that crystallization can start at approximately 773 K \citep{Jaeger:03}. This low crystallization temperature has been observed in silicates that contain internal hydroxyl groups. The heating process of the silicate dust grains in the protoplanetary disk may have induced the crystallization process and the thermal desorption of chemically bound water at temperatures ranging from 773 to 873\,K. It is important to note that the timescales for heating and processing of silicate dust grains in protoplanetary disks are significantly longer than those employed in a laboratory setting. Consequently, the temperatures required for desorption could be lower than the values determined in the laboratory. 
Determining the precise heating rate for the annealing of dust components in protoplanetary disks is challenging. Assuming a radial drift through a more static disk, the radial-velocity change per unit time is calculated to be $dT$/$dt$ = 10$^{-3}$ -- 1\,K\,yr$^{-1}$.  However, transient heating events, such as accretion bursts, have the potential to result in temporary increases in the rate of heating.

As already explained, it is possible to derive the activation energy from thermogravimetric measurements or other temperature-programmed experiments measured at different heating rates by using the Kissinger equation. 
The determination of the activation energy peak positions facilitates the estimation of the relevant peak temperatures of desoprtion for different annealing rates. The peak values derived at three distinct heating rates were found to correspond to an activation energy of 273\,kJ\,mol$^{-1}$. Assuming an annealing rate \textit{dT/dt}=10$^{-3}$\,K\,yr$^{-1}$, the high-temperature desorption was estimated to be approximately 654\,K. However, it should be noted that such an extrapolation over a few orders of magnitude is only a very rough approach, and the error of the derived temperature for astrophysical annealing rates is at least $\pm$40\,K.

In addition to water-containing amorphous silicates, crystalline phyllosillicates can be essential tracers of water in disks. The amount of water in phyllosilicates can reach about 3 \% of the total mass of refractory material and the timescale for their formation is shorter than the lifetime of protoplanetary disks \citep{Thi:20}. Phyllosilicates occur as a major component in carbonaceous chondrites and interplanetary dust particles in a significant fraction \citep{Morris:76,Nakamura:05}. 
Furthermore, the presence of phyllosilicates has been identified in returned samples from the asteroid Bennu and from the carbonaceous asteroid Ryugu by the OSIRIS-REx and the Hayabusa2 missions, respectively. \citet{Sandford:25} has demonstrated that the measured IR spectra of the particles returned from Bennu are dominated by phyllosilicate-carbonate mixtures. The analysis of the returned sample from Ryugu has revealed the presence of phyllosilicates and saponite, which is another layer silicate mineral that contains hydroxyl groups \citep{Leroux:24}.
Phyllosilicates are typically formed through the early parent body process of aqueous alteration, often associated with the co-precipitation of carbonates and phosphates. 

Carbonaceous chondrites are a group of meteorites and among them the CI and CM chondrites are the most primitive solar system material, which are strongly affected by aqueous alteration at rather low temperatures ($<$ 323\,K for Hayabusa samples). Moreover, the presence of presolar grains and phyllosilicates has been identified in both types of chondrites (Morris, 1976). Even more compelling, the presence of 'amorphous serpentine' has been identified in meteoritic samples. \citet{Nakamura:05} performed experiments on the annealing of various meteorites and discussed the dehydration of meteoritic phyllosilicates at different temperatures. The author reported that heating led to a dehydration of phyllosilicates, resulting in the formation of crystalline olivine at temperatures ranging from approximately 773 to 1023\,K, precisely the same temperatures measured for the desorption process of serpentine minerals. 

\citet{Perotti:23} found that the observed water spectrum in the 6.78--7.36 $\mu$m spectral region is best fitted with a slab of gas at a temperature of 600\,K, with an emitting area of radius R = 0.047\,au and a column density of \textit{N} = 1.4 $\times$ 10$^{18}$ cm$^{-2}$. It is possible to perform a rough estimate of the mass of water molecules by using the equation
\textit{M$_{H_2O}$} = $\pi R^2$ $\times$ $N_{H_2O}$ $\times$ $m_{H_2O}$. The resulting water mass was estimated with $M_{H_2O}$ = 6.5 $\times$ 10$^{19}$ g or $M_{H_2O}$ = 1.1 $\times$ 10$^{-8}$ $M_\Terra$. \citet{Benisty:21} derived the inner dust mass of PDS 70 from ALMA observations, yielding  a range of 0.08 -- 0.36 $M_\Terra$. Assuming that the whole dust mass of silicates exhibits a water content between 0.1 -- 5\,wt\%, a potential water reservoir of 10$^{-4}$ -- 2 $\times$ 10$^{-2}$\,$M_\Terra$ can be expected. Consequently, a very low fraction of much less than 1\,\% of the hydrated silicates is required to produce the observed column density of water molecules.
  
However, it is imperative to note that the characteristic temperatures measured in thermogravimetric analy\-ses are contingent on the thermal history of the amorphous silicates \citep{Zhuravlev:01}. This extends to other properties as well, including the viscosity behavior of amorphous silicates and the crystallization temperatures \citep{Vogel:92,Jaeger:03}. Given the absence of a comprehensive knowledge regarding these sample parameters, it is necessary to consider temperature ranges rather than specific values. 

The findings of our experimental study support an alternative hypothesis, where water could be released by a thermal dehydration process of silicate materials with varying compositions that contain chemically bound water.  The derived dust temperatures in PDS\,70 between 400 and 600\,K are supportive to the thermal desorption scenario as the origin of the gaseous water reservoir. The presence of crystalline silicates serves as an additional indication of the heating processes that have taken place. 

\section{Conclusions}

An experimental study was conducted to investigate the desorption characteristics of various magnesium silicate materials containing different amounts of physically and chemically adsorbed water. To this end, two laboratory silicate dust analogs and natural minerals were subjected to thermogravimetric analyses within a temperature range of 293 to 1273\,K. The experimental findings offer a supplementary rationale for the provenance of water molecules within the warm zone of the inner disk of PDS\,70, as well as for other water gas reservoirs located in the inner regions of protoplanetary disks. The desorption of water molecules from water containing silicates can be achieved through thermal processes. Water that is physically and chemically bound exhibits a distinct desorption behavior. 
Physically bound water desorbs up to a temperature of 473\,K, while hydroxyl groups bound to Si or Mg in amorphous silicates or in minerals desorb at temperature around 773\,K at annealing rates of about 20\,K. It has been demonstrated that under astrophysical annealing rates, the desorption temperature can be considerably decreased. For the Erbendorf Serpentinite sample, the derived desorption temperature for an annealing rate of approximately 10$^{-6}$K\,min.$^{-1}$ was found to be around 650\,K. The presence of hydroxyl groups in interstellar silicates is likely attributable to the interaction of dust and water ice in diverse interstellar environments under varied conditions. The experimental results support the hypothesis that water molecules are desorbed from silicate grains of different composition and structure that contain chemically bound water in the form of various hydroxyl groups upon heating in the inner disk of PDS\,70. The water would be released at much higher temperature than the snowline temperature where water ice sublimates.

\section{Acknowledgments}
We wish to express our profound gratitude to Renzo Paulus, who carried out the thermogravimetric measurements at the Institute of Organic Chemistry and Makromolecular Chemistry (IOMC) at the Friedrich Schiller University Jena. We are also grateful to Michael Achtellik for the performance of DEGAS measurements. In addition, we would like to thank Giulia Perotti for her valuable discussions.

\bibliographystyle{aasjournal}


\end{document}